# Physical Characteristics of Umbral Dots Derived from a High Resolution Observation

Ali Kilcik[1] · Volkan Sarp[1] · Vasyl Yurchyshyn[2] · Jean-Pierre Rozelot[3] · Atila Ozguc[4]

[1] Department of Space Science and Technologies, Akdeniz University Faculty of Science, 07058 Antalya, Turkey

[2] Big Bear Solar Observatory, Big Bear City, CA 92314, USA

[3] Université Côte d'Azur (UCA), 06130, Grasse, France

[4] Kandilli Observatory and Earthquake Research Institute, Bogazici University, 34684 Istanbul, Turkey

**Abstract**
The aim of this study is revisit the physical parameters of umbral dots (UDs) with the latest high resolution observations and contribute to the scientific understanding of their formation and evolution. In this study, we applied a particle tracking algorithm for detecting UDs in NOAA AR12384 observed on June 14, 2015 by the Goode Solar Telescope (GST). We analyzed average position distributions, location dependencies, and general properties of detected total 2892 UDs separately during their life time and the periodic behavior of only selected 10 long living UDs. We found; i) brightest, largest, fastest and most elliptic UDs tend to be located at the umbra-penumbra boundary while their lifetime does not display any meaningful location dependency, ii) average dynamic velocity of all detected UDs is about twice (0.76 km/s) of the previously reported average values, iii) obtained trajectories from the longest living 354 UDs show that they have generally inward motion, iv) chosen 10 long living UDs generally have similar periodic behavior showing 8.5–32, 3.5–4.1, 1.5–1.9, and 1.1–1.3 minutes periodicities, v) generally, detected UDs have an elliptical shape with the averaged eccentricity of 0.29, with a 0.11 standard deviation, vi) larger UDs tend to be more elliptic and more dynamic.

## 1. Introduction

A developed sunspot consists of two main regions which are a dark central part called umbra and a lighter part surrounding the umbra called penumbra. A typical sunspot umbra consists of various transient and fine structured features, which are manifestations of magnetic activity inside the umbra (e.g., Yadav, Louis, and Mathew, 2018) Among these: fine-scale structures are umbral dots (UDs) that were first observed in 1916 by Stanislas Chevalier as bright specks in the umbra (Choudhuri, 1986). It has been long known that sunspot brightness is lower than the solar disk since their strong magnetic field suppresses vertical convective motions and inhibits the energy transport from the interior layers. Although UDs only cover 3% to 10% of total umbral area, they contribute up to 20% of the total umbral brightness (Watanabe et al., 2012). Understanding the physical processes of UDs emergence is expected to shed light through more realistic sunspot models.

There are two alternative models proposed to explain sunspots and the origin of UDs: The first one, known as a cluster model, suggests that the sunspot magnetic field beneath the visible surface is composed of many individual flux tubes loosely clustered together and UDs are the apex of field free material penetrating through the surface between these flux tubes (Parker, 1979; Choudhuri, 1986). The other one, known as a monolithic model, is based on the magnetoconvection mechanism in a single large block of flux tube that constitutes the sunspot (Weiss, Proctor, and Brownjohn, 2002). The monolithic model is supported by 3D MHD simulations that explain the appearance of UDS as buoyant rise of narrow plumes that govern the convective energy transport (Schüssler and Vögler, 2006).

It has been challenging to analyze the structure of UDs due to their small size close to the resolution limits of telescopes and strong variations of image quality in earlier years (e.g. Beckers and Schroter, 1968). Sobotka, Brandt, and Simon (1997) showed that UD parameters are highly dependent on the observational and methodological conditions, so they do not have a typical size or lifetime. However, basic dynamics of UDs came into focus, such as their spatial distributions in the umbra with the improving observations. Grossmann-Doerth, Schmidt, and Schroeter (1986) distinguished two distinct classes of UDs based on their locations in umbra. Thus central UDs (CUDs) are located well within the umbra and were demonstrated not to exceed the minimum umbral intensity level by more than two orders of magnitude. On the other hand, peripheral UDs (PUDs) are located close to the umbra-penumbra boundary and their intensities are above the umbra average.

Typical characteristics such as size, velocity, filling–factor, and lifetime of UDs were studied intensely, owing to the advancing observational instruments, increasing resolutions and image processing techniques. Analyzing 1–m Swedish Solar Telescope data, Sobotka and Hanslmeier (2005) found that the majority of the UDs have a diameter around 0".23 which corresponds to 170 km on the solar surface. Riethmüller et al. (2008), used the same telescope to show that the histogram of UDs lifetimes follows an exponential distribution, i.e., UDs do not have a typical lifetime. These authors also found that the mean UD diameter varies between 50 and 750 km.

Later, Hinode Solar Optical Telescope data were used by (Watanabe, Kitai, and Ichimoto, 2009) to reveal that the mean lifetime (25–34 minutes) and diameter (∼ 190 km) of UDs are not changing with varying magnetic field strength of sunspots. Watanabe et al. (2010) analyzed one particularly fast (1.3 km/s) UD using Dunn Solar Telescope data and found its lifetime and diameter to be 8.7 minutes and 240 km, respectively. Kilcik et al. (2012) used data recorded by the Goode Solar Telescope (GST) and showed that UD diameters vary from 0".23 to 0".41 (166–298 km), and none of these fine scale structures has an exact circular shape. These authors also found an anti–correlation between the lifetime and velocity of UDs. Feng et al. (2015) studied similarities and distinctions between PUDs and CUDs analyzing data from Hinode Solar Optical Telescope. Their results show that sizes of PUDs and CUDs are not so distinctive: 224±65 and 228 ±67 km, respectively. These authors also calculated the eccentricity parameter of UDs to explore their deviations from circular shape and confirmed the previous results. Finally, Yadav, Louis, and Mathew (2018) studied physical properties of UDs observed in different sized sunspots but could not find any significant relationship between the investigated physical parameters of UDs and those of sunspots (area, epoch, and decay rate).

It is clear that some of the physical parameters of UDs can be measured more accurately with the increasing resolution of observations. However, there are still debates regarding other parameters such as the lifetime. This study presents a detailed analysis of UDs, which have the potential to shed light onto the energy balance mechanism of sunspots required to fully understand their structures and Dynamics. Plan of the paper is as follows. In Section 2, we present the data used and the method of analysis. Results are presented in Section 3, including the characteristic parameters, whereas final Section 4 ends with conclusions and discussions.

## 2. Data and Methods
### 2.1. Data Preparation

We studied statistical properties of UDs detected in NOAA AR 12384, located at 171",–352" of heliographic coordinates observed with a TiO broadband filter (red continuum) installed on GST, which was stabilized by a high-order adaptive optics (AO) system (Shumko et al., 2014). The AR was observed from 16:30 to 18:08 UT on 14 July 2015. The data set is nearly continuous except for a few very small gaps, which we did not take into account since we did not address any time dependence. The TiO filter has a 10 Å bandwidth that is centered at the wavelength of 7057 Å. This spectral line is sensitive to temperature, and it is exceptionally suitable for observing dark and cool regions such as sunspot umbra and penumbra

(Berdyugina, Solanki, and Frutiger, 2003; Riethmüller et al., 2008; Abramenko et al., 2010). The TiO data were corrected for dark currents and flat fielded. The Kiepenheuer-Institute Speckle Interferometry Package (Wöger and von der Lühe, 2007) was applied to produce speckle–reconstructed images with a field of view (FOV) of 70"x 70". The pixel scale was 0.0342. The cadence of the TiO data was 15 s. In Figure 1 we show temporal evolution of the investigated AR for the observation period.

The data set was further processed as follows: (1) all images were co-aligned and the area outside of the umbra was removed from the image by using a binary filter (Figure 1, left panel); (2) the brightness of each image was adjusted to the average level of the set; and (3) band pass filters were applied to reduce the noise level in the masked images. The band pass filter utilizes a wavelet technique based on convolution with a "Mexican hat" kernel to remove random digitization and the background noise (Crocker and Grier, 1996). Here, we applied a spatial bandpass filter to smooth sunspot images and subtract the noisy background. The spatial wavelength lower and upper cutoffs for the filter were 1 and 7 pixels, respectively. The spatial wavelength lower cutoff value was chosen to be the smallest possible value and the upper cutoff value was determined after multiple runs of the algorithm as that produced the best detection results in terms of accuracy and precision. The resulting data cube had a FOV of 450 x 450 pixels (15.39 x 15.39) (Figure 2, left panel).

### 2.2. UD Tracking

We used the IDL particle tracking code by Crocker and Hoffman (2007), originally written for blood cell detection and tracking. We took a square root of the calculated radius as mentioned in the web page[1]. The outcome of the algorithm performance on a single frame is shown in the middle panel of (Figure 2).

In the original code, the upper cutoff parameter for the bandpass filter is the diameter of the particles that will be searched. Because of the varying size of UDs, seeing, and image pre-processing, we modified the original algorithm to use the diameter search parameter nearly two times larger than the upper cutoff parameter that ensured accurate detection performance. The minimum allowed value for the peak brightness was determined by checking the maximum lifetime of UDs. Although there is no direct connection between them, selecting very low minimum allowed peak brightness resulted in detecting noise features. So this value was determined to limit the maximum lifetime of detected UDs to be less than the total observation time. Thus, we eliminated some features which have a lifetime equal to the total observation time (97.5 min). This allowed us to eliminate all indistinct noise features that could have been introduced by instrumentation, seeing, processing, etc.

---

[1] http://www.physics.emory.edu/faculty/weeks//idl/tracking.html

This algorithm detected a total of 2892 UDs, all of which were tracked on the duration of their lifetime. If a tracked UD was detected to jump more than 5 pixels (~ 124 km) between consecutive frames, it was considered as a new event after the jump. Also, any UD that cannot be tracked in at least 3 consecutive frames was considered to be noise and was discarded.

Five characteristics of UDs, namely, brightness, calculated as the sum of pixel brightness inside of the each UD area, diameter, eccentricity, life- time, linear velocity and dynamic velocity were calculated. The normalized brightness, diameter, and the eccentricity were calculated for each image in the time series, and an average for each frame was then determined. The lifetime was calculated from the total number of frames in which a given UD was tracked. The linear velocity was calculated by dividing the direct distance between the fade-out and emergence locations of a UD by the lifetime of the same UD. To calculate dynamic velocities of UDs, we measured the length of the displacement vector of a UD in two consecutive frames and then calculated the total length of these displacement vectors determined from each pair of consecutive images in the time series. The total length was then divided by the lifetime to obtain the dynamic velocity of each UD. To investigate the movement of UDs inside the umbra we chose 354 long living (longer than 15 minutes) UDs and tracked them during their lifetime. Then we plotted the central coordinates of these UDs to visualize their trajectories.

### 2.3. Brightness Oscillations

To investigate the periodic behavior of UDs we chose 10 UDs that had the longest lifetime (about one hour) and were distributed nearly homogeneously over the umbra (see Figure 1, right panel, white circles). Their brightness variations were analyzed by using the Multi Taper Method (MTM) that allows us to obtain possible periodicities with a high degree of statistical significance over the lifetime of each of the selected UDs. The method uses orthogonal windows (or tapers) to obtain an estimate of the power spectrum (for more details see Thomson, 1982; Ghil et al., 2002). Here we used three sinusoidal tapers, and the frequency range was limited to 0.031 (32 minutes) –1 (1 minute). The significance test was carried out assuming that the noise has a red spectrum, and a signal was detected when the 95 % confidence level was reached. To show MTM power spectra for all selected UDs in the same graph, the power spectrum of each UD was normalized by dividing the power by the 95 % confidence level value of the same UD. In other words, to normalize the power spectrum we divided power at each frequency by the 95 % confidence value of the same frequency.

3. Results

As the first step of our analysis, we plotted histograms of all UD parameters (Figure 3). The bar plots of the brightness, lifetime and linear velocity are not symmetrical, and show a heavy tail on large scales, which is indicative of a lognormal distribution. The histograms of the dynamic velocity, eccentricity and the diameter are nearly symmetrical with the dynamic velocity and eccentricity distributions having only a slightly extended tail at higher values, while the diameter shows a very small tail at smaller ranges. The maximum, minimum and mean values of these parameters are presented in Table 1. As shown in the brightness histogram (upper left panel in Figure 3) there is a secondary peak located around $4.0 \times 10^5$ that possibly comes from faster and more eccentric UDs. Their average values are respectively calculated as 1.05 km/s and 0.37 for the brightness interval between $3.0 \times 10^5$ and $5.0 \times 10^5$, which correspond to the secondary peak in the brightness histogram.

The relationship between these parameters was then analyzed using regression and correlation methods. To estimate the 95 % confidence intervals of correlation coefficients (CC), we applied Fisher test, which gave us upper and lower bounds of the confidence level for the obtained correlation coefficient. Four strongest relationships, in terms of correlation coefficients, out of ten possible combinations of all parameters are shown in Figure 4. The brightness shows nearly the same level of correlation with both diameter and dynamic velocity (CC=0.43±0.03, CC=0.40±0.03, and CC=0.38±0.04, respectively). The UD diameter exhibits a much higher correlation with eccentricity (CC=0.53±0.03). Generally speaking the relationship be- tween the above mentioned parameters is better pronounced at a smaller data range, while at larger scales there is little or no dependence. The exception is the relationship between the diameter and the eccentricity, where larger UDs are more eccentric.

To analyze the location dependencies of these parameters, central coordinates of each UD were averaged during their lifetime and plotted against their mean parameters (Figure 5). We found that the lifetime does not display any meaningful location dependency and thus it was not taken into account and plotted. As expected, we found that brightest UDs tend to locate at the umbra-penumbra boundary and thus they can be considered as Peripheral UDs (PUDs). Larger and the most elliptic UDs also tend to concentrate at the umbra-penumbra boundary, although the location dependence for size and eccentricity is not as strong as those for the brightness. There is a slight preference for dynamic UDs to locate at the boundaries, but UDs with high dynamic velocities can also be found at the central part of sunspot umbra (right panel of (Figure 5)).

We tracked the movement of long lived (longer than 15 minutes) UDs during their lifetime. Total 354 UDs out of 2892 have been selected and tracked. The obtained trajectories show that the UDs generally display inward motion, although there are a few UDs that either move toward the periphery or follow a non-uniform, chaotic trajectory located in a small area (see, Figure 6).

We also investigated the spatial variations of these parameters from the umbral center to umbra-penumbra boundary for selected 354 UDs (See, Figure 7). Both brightness and dynamic velocity plots do not show any variation up to some distance (about 2500 km) from the umbral center after which they rapidly increase. This pattern is also present in the other two plots (diameter and eccentricity) although it is less pronounced due to larger data scatter.

For a detailed investigation of the variations shown in Figure 7, we selected 10 UDs located at the outer bound of umbra (as shown in Figure 2 right panel, red and blue marks). As the first step, we plot their distance to the center of the umbra during their lifetime (Figure 8, upper left panel). The graph demonstrates that all selected UDs tend to move to the umbral center as time progresses. Their brightness, radius and eccentricity also tend to decrease as they move closer to the center, as is also shown in Figure 6. However, there is a well pronounced peak in the radius and eccentricity plots centered at around 2000 km, and the variation pattern for these parameters change after the distance from the umbra center exceeds approximately 2500 km. The brightness shows a similar pattern with the change point occurring at nearly 3000 km. This change in the spatial dependency of the UD parameters suggest the existence of two distinct populations of UDs, namely PUDs and CUDs.

As a final step we have chosen 10 UDs (Figure 2 right panel, marked with white and blue) that had life–time exceeding 45 minutes and investigated their brightness fluctuations. We found that all these selected UDs show about a similar periodic behavior with at least 95 % confidence level. These periodicities are 8.5–32, 3.5–4.1, 1.5–1.9, and 1.1–1.3 minutes, respectively. The 8.5–32 minutes periodicity is manifested in all analyzed UDs, the 1.5–1.9 minutes periodicity exists in six of them (UD1, UD2, UD4, UD5, UD6, and UD8). The 1.1–1.3 minutes were found for only half of the long lived UDs (UD1, UD3, UD4, UD8, and UD10) while 3.5–4.1 periodicity found only in three UDs (UD1, UD4, and UD8) (see, Figure 9).

## 4. Conclusions and Discussions

In this study, we analyzed the distribution, location dependencies, and general properties of 2892 UDs and brightness oscillations for selected long living UDs (total 10) automatically detected in the umbra of the main spot of NOAA AR 12384 observed by the GST on June 14, 2015. Our findings are as follows;

- The average velocity of the analyzed UDs is about twice (0.76 km/s) of the previously reported values (approx. 0.4 km/s) in the available literature.
- Profiles of the brightness and dynamic velocity variations with the distance from the center of the umbra can be used as a discriminator between the CUDs and PUDs. The brightest, largest, fastest and most elliptic UDs show preference to locate at the umbra-penumbra boundary, while the lifetime does not display any meaningful location dependency.
- Tracking of UDs during their lifetime shows that they generally move inward toward the center of the umbra.
- Intensities of all longest living UDs generally show a similar periodic be- havior with periods of 8.5–32, 3.5–4.1, 1.5–1.9 and 1.1–1.3 minutes.
- Most of the analyzed UDs have not a circular shape. The average eccentricity is 0.29, with a 0.11 standard deviation.
- Larger UDs tend to be more elliptical and more dynamic (larger dynamic velocities).

Many studies found the horizontal velocity of UDs to be about 300-400 m/s (Rimmele, T. and Marino, J., 2006; Riethmüller et al., 2008; Watanabe, Kitai, and Ichimoto, 2009; Kilcik et al., 2012, and references therein). Watanabe, Kitai, and Ichimoto (2009) calculated the velocity amplitude of observed UDs by dividing the direct distance between the fade-out and emergence locations of a UD by the lifetime of the same UD. They obtained the average velocity of 0.44 km/s based on 2268 UD measurements. Here, we used the total length of the path traveled by a UD and obtained the average UD velocity of about 0.76 km/s, which is nearly twice of the previously reported values with 50 m/s minimum and 3.84 km/s maximum values (Watanabe, Kitai, and Ichimoto, 2009; Feng et al., 2015). This is the first time that such a high average horizontal velocity is obtained for the UDs observed in a sunspot umbra. We found that this difference comes from the method used to calculate the UD velocity. Tracking UD trajectories indicates that UDs may not be traveling in a straight line but rather wandering as they move closer to the umbral center. Thus, we may suggest that to calculate more accurate UD velocity all motions that UD does during it life time should be taken into account.

UDs are generally separated into two types depending on their location within central and peripheral (Riethmüller et al., 2008; Watanabe, Kitai, and Ichimoto, 2009; Watanabe et al., 2012; Feng et al., 2015, and references therein). On average, peripheral UDs are larger, brighter, and move faster as compared to central ones (Riethmüller et al., 2008; Kilcik et al., 2012, and reference therein). Here we considered a continuous spatial distribution of these parameters (brightness, diameter, eccentricity and velocity) and found that the largest, most elliptic, brightest, and fastest UDs tend to locate at the boundaries of umbra-penumbra (see Figure 5). The average brightness and dynamic velocity vary with the distance from the center of the umbra and these parameters rapidly increase at the distance about 2500 km from the center. These findings are consistent with previous reports and further suggest that this distance can be used as a separation distance for CUD and PUDs.

It was noted that the central UDs are nearly static, while the peripheral UDs move toward the center of the umbra with speeds less than 1.0 km/s (Sobotka, Brandt, and Simon, 1997). Later, Watanabe, Kitai, and Ichimoto (2009), Louis et al. (2012), and Feng et al. (2015) confirmed these inferences. Our findings based on tracking 354 UDs which were living longer than 15 minutes, confirmed previous results and further showed that most of the longest living UDs moving toward the umbral center.

Sobotka, Brandt, and Simon (1997) reported that the average UD size increases with the lifetime of a UD. A positive relationship between the brightness and diameter of UDs had previously been reported by Tritschler and Schmidt (2002). Riethmüller et al. (2008) reported that the brighter UDs are on average slightly bigger than the dimmer ones. Later, Kilcik et al. (2012) concluded that UD velocities show anti-correlation with lifetime and UD diameters increase with the brightness only in the brighter umbral areas. In this study, we found that UD brightness shows about the same level of correlation with both diameter and dynamic velocity .The dynamic velocity shows nearly the same level of correlation with the UD diameter and a much higher correlation with the UD eccentricity. We also found that the relationship between some UD parameters (except diameter and eccentricity) is better pronounced at smaller data ranges and the dependence is nearly absent at large data ranges. Note that we could not find any meaningful relationship between the lifetime and dynamic velocity as reported previously.

The convective energy transport is dominated by upflow plumes whose top parts are forced by the magnetic field to loose buoyancy and shape into cusp like structures. The resulting bright features are UDs (Schüssler and Vögler, 2006). Intersection of plumes with the umbra may cause the slightly elongated shapes of UDs. Although these upflow plumes are nearly

field–free, MHD simulations show their coherent settlements along the magnetic field lines (Rempel, Schüssler, and Knölker, 2009), which are long known to have an increasing inclination through to the penumbra. This phenomenon is also supported by our results (See lower left panel of Figure 7). UDs close to the central part of umbra have a smaller mean eccentricity (0.27) compared to outer part where the eccentricity is about 10 % higher (0.30). On the contrary, flux tubes are shown to not being able to reproduce the UDs although the convection mode is similar to the intrusion of unmagnetized plasma in a shallow cluster model (Rempel and Schlichenmaier, 2011). Our results are in agreement with their results that UDs are a result of magnetoconvection in the form of upflow plumes at a monolithic sunspot model. Thus we may suggest that elongated shapes of UDs are a natural result of the inclination of plumes.

Size of UDs were studied intensely in the past (Sobotka and Hanslmeier, 2005; Riethmüller et al., 2008; Watanabe et al., 2012; Kilcik et al., 2012; Feng et al., 2015; Yadav, Louis, and Mathew, 2018, and reference therein) and they reported the UD diameter to be between 50–750 km. Feng et al. (2015) analyzed size of Hinode/SOT UDs depending on their location (peripheral and central) found that the size of these two types of UDs are nearly the same (about 225 km). Here we analyzed much larger sample (2892 UDs) and reported that the diameter of the UDs vary from 107 to 276 km with the average value of 207 km. Our findings show a very good agreement with the diameter values found in the literature.

Schüssler and Vögler (2006) found that simulated UDs have horizontally elongated shapes using their numerical simulations. Kilcik et al. (2012) concluded that all observed UDs without exception have an elongated shape. Feng et al. (2015) also reported an eccentric shape of UDs with the average eccentricity of 0.75. We found in this study an average eccentricity value (0.29) much smaller than their result. The difference probably comes from the difference in resolution of data used and how the UDs were identified and measured. Also the main sunspot of NOAA AR 12384 in our study is more symmetric and larger than that used in the study of Feng et al. (2015). Note that the AR we studied was very stable and did not show any serious change during the observation period (see, Figure 1). We think that the morphology and evolution of the umbra may also affect the results.

Using data obtained by CRISP imaging spectropolarimeter at the Swedish 1 m Solar Telescope Watanabe et al. (2012) reported that the average UD lifetime to be nearly 18 minutes. They concluded that this high lifetime was possibly due to the manual detection procedure that they used, which was capable of detecting fainter UDs. Kilcik et al. (2012) analyzed UDs using GST data and found that the average lifetime is about 8 minutes and varies between 2.5 and 34.5 minutes. Later, based on Hinode data Feng et al. (2015) reported that the UD life time varies from 1 to 36 minutes. The lifetime of UDs in this study varies from 0.75 to 65 minutes with the average value of 6.92 minutes, which is in very good agreement with previous measurements.

The periodicity analysis of the UDs light curves was first carried out by Sobotka, Brandt, and Simon (1997). They analyzed intensity light curve of the five longest living (longer than 126 minutes) UDs observed in NOAA AR7519 with the Swedish Vacuum Tower Telescope. They concluded that UDs have periodicities of 32, 16, 11, 8.5, 7.2 and 6.1 minutes. Later, Watanabe, Kitai, and Ichimoto (2009) observed UDs in a stable sunspot (NOAA AR10944) by the Hinode Solar Optical Telescope (SOT) on March 1, 2007. They also analyzed the light curve of 76 UDs living longer than 30 minutes and reported 8–16 minutes oscillations. Recently, Ebadi, Abbasvand, and Pourjavadi (2017) analyzed intensity oscillation of 17 UDs observed in NOAA AR10944 with SOT/Hinode on March 1, 2007. They used Morlet Wavelet analysis method for the analysis and concluded that UDs oscillate in two separate period ranges of 180–400s (3– 6.7 minute) and 450–720s (7.5–12 minutes). Yuan et al. (2014) concluded that the oscillation period increases with the increasing distance from the sunspot center and explained this phenomenon with the inclination of magnetic field lines. Here, we applied MTM period analysis method to 10 long living UDs (longer than 45 minutes) observed with GST and detected four different period ranges (8.5–32, 3.5–4.1, 1.5–1.9 and 1.1–1.3 minutes). Thus, we confirm above results and found further two period ranges which are shorter than previously reported, for the UD oscillations. The 1–2 minutes periodicities found in this study confirms Yuan et al. (2014)'s results and could explain the high frequency oscillations of umbral dots. These short period ranges may contribute to well understanding of sunspot umbral oscillation and may contribute to theoretical studies from the point of background physical mechanisms of UDs.

**Acknowledgments** The high resolution sunspot data are taken from BBSO/GST. BSO operation is supported by NJIT and US NSF AGS-1821294 grants. GST operation is partly supported by the Korea Astronomy and Space Science Institute (KASI), Seoul National University, and by strategic priority research program of CAS with grant No. XDB09000000. This study was supported by Project 117F145 awarded by the Scientific and Technological Research Council of Turkey. V.Yu. acknowledges support from AFOSR FA9550-19-1-0040, NSF AGS-1821294, AST-1614457, and NASA HGC 80NSSC17K0016 and GI 80NSSC19K0257 grants.

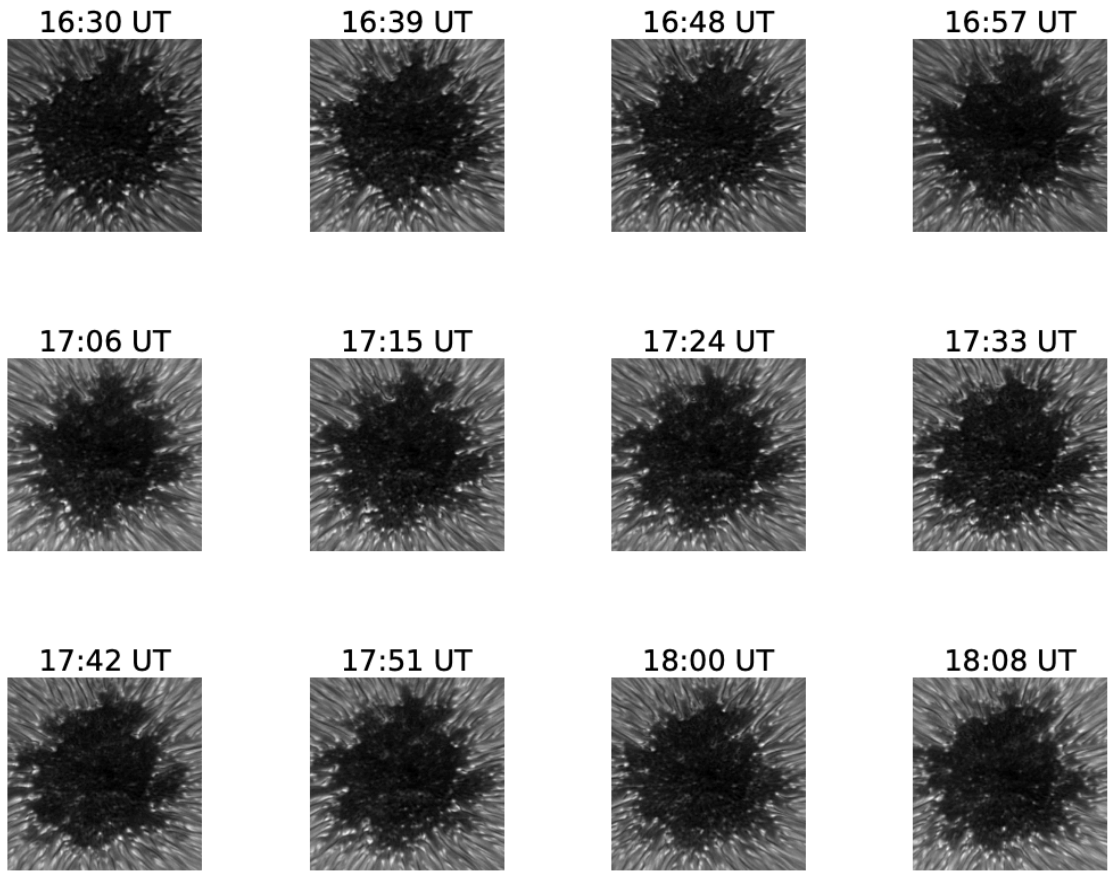

**Figure 1.** Temporal evolution of the NOAA AR12384 for the observation period. Times in the top of each figure show the data receiving time. The FOV of all images are 450 x 450 pixels (15.39 x 15.39)

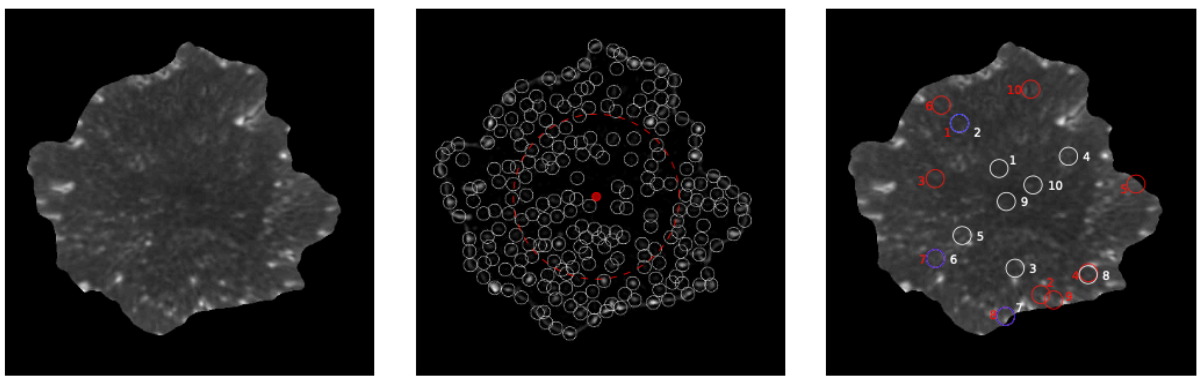

**Figure 2.** Separated umbra of the main sunspot in NOAA AR 12384 (left panel), all detected UDs (circled) in one sample image (middle panel), where the red circle separates periphery and central UDs. In the right panel we indicate with white circles UDs that were used in the oscillation analyses, red circles mark UDs that were used in variation analysis, and blue marked UDs were used in both types of analysis. FOV of these images are 450 x 450 pixels

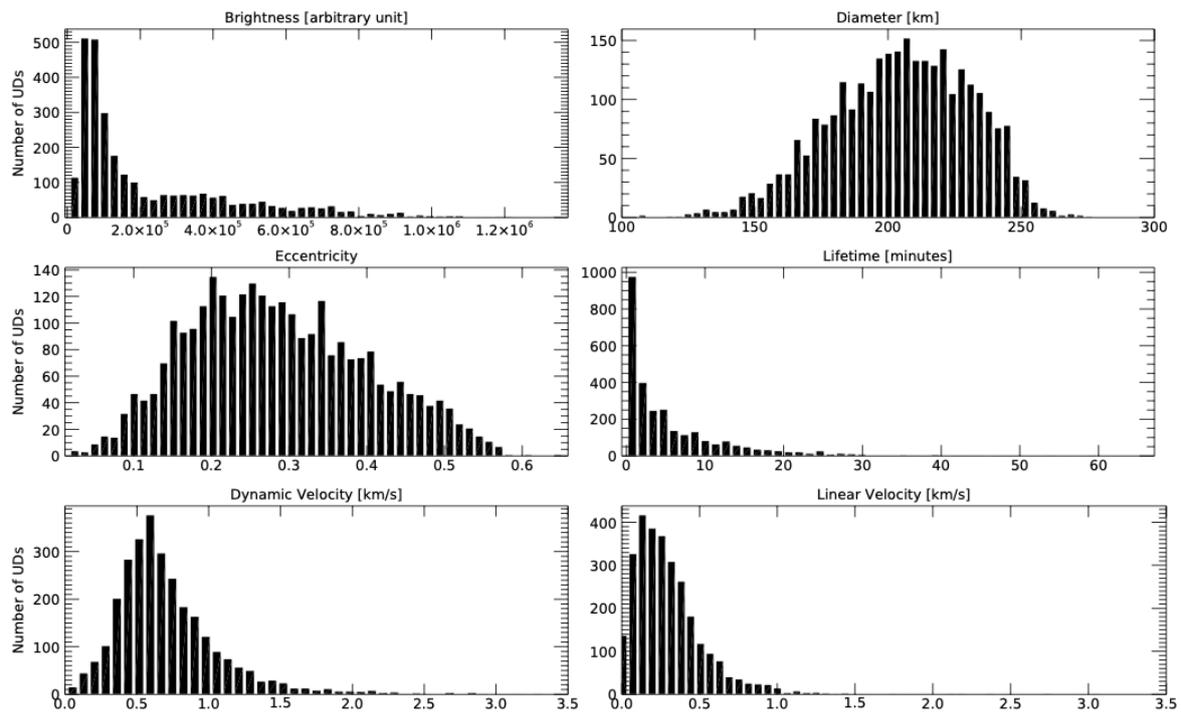

**Figure 3**. The distribution of average characteristic parameters of all detected UDs for the investigated AR.

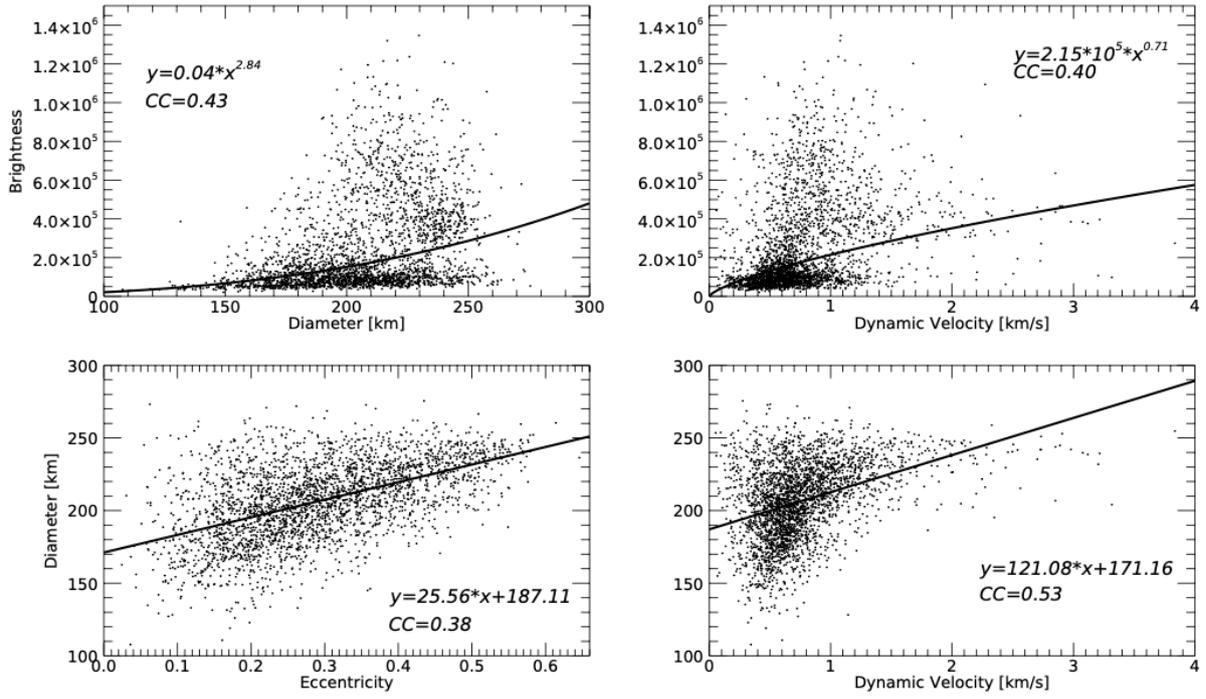

**Figure 4.** Relations between characteristic UD parameters for NOAA AR 12384.

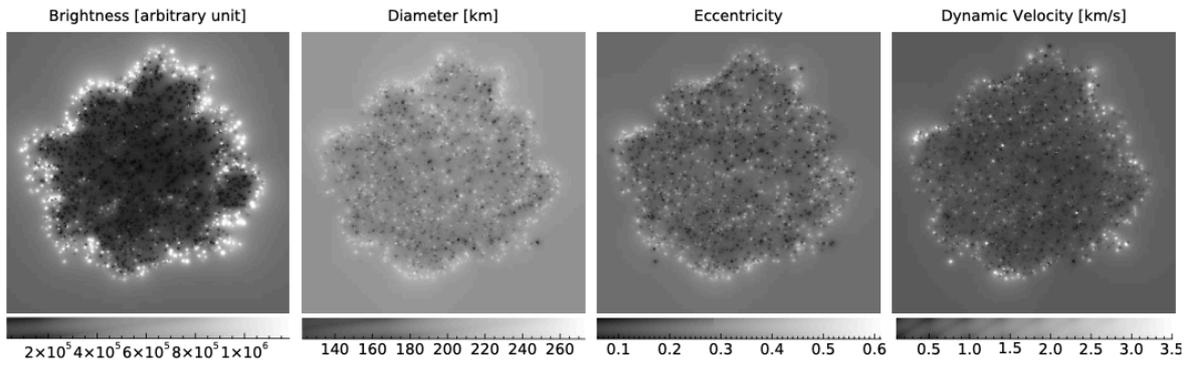

**Figure 5.** Location dependencies of four UD parameters for NOAA AR12384. The FOV is 450x450 pixels.

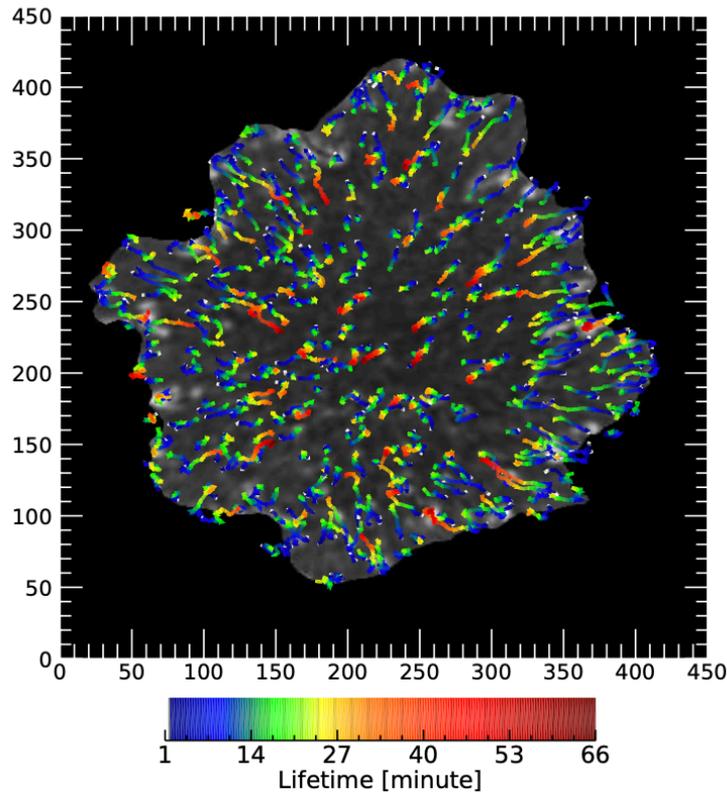

**Figure 6.** Trajectories of selected long living (longer than 15 minutes) all UDs (354 UDs). The FOV is 450x450 pixels.

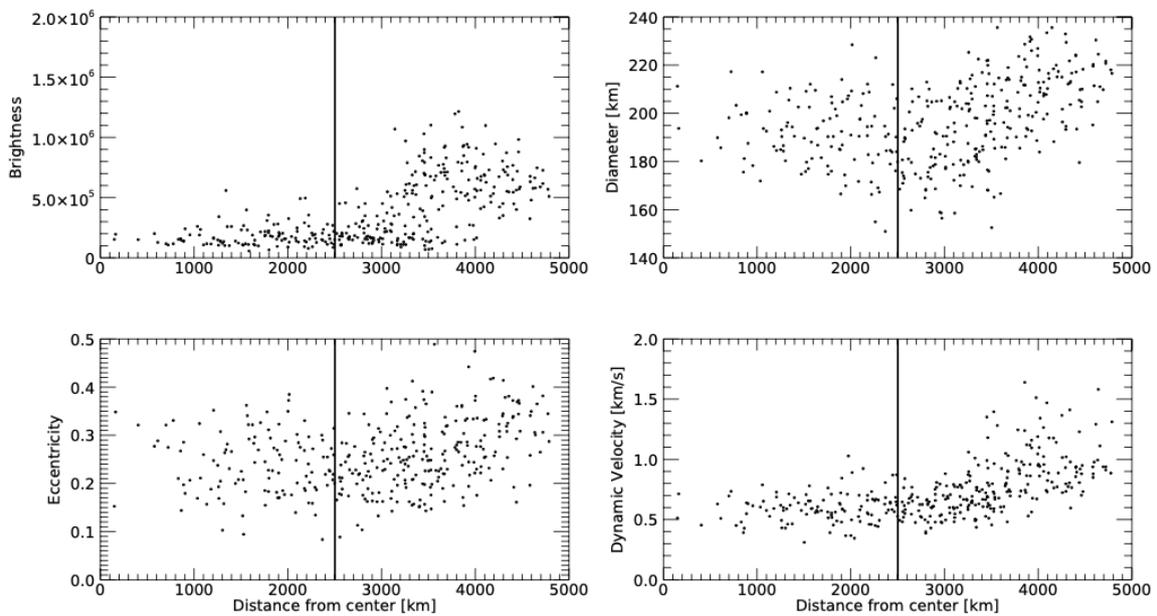

**Figure 7.** Variations of UD parameters plotted for 354 long living (longer than 15 minutes) UDs with the distance from the umbral center. The vertical line which shows the brightness change starting distance, represents the boundary between CUDs and PUDs.

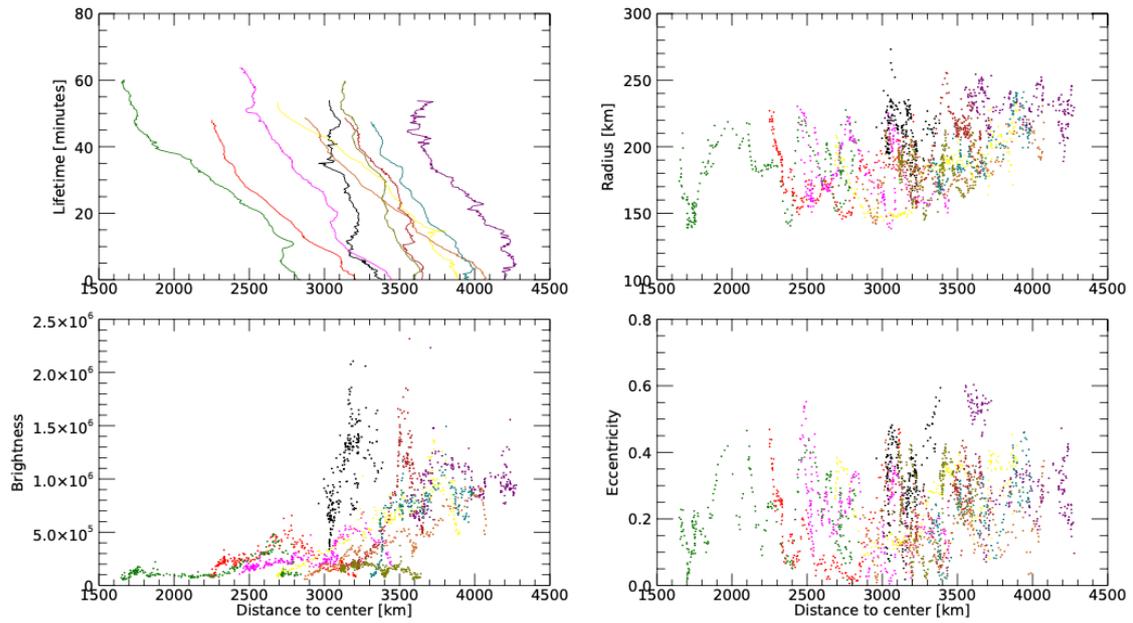

**Figure 8.** Variations of 10 selected UDs according to their lifetime and distance to the center.

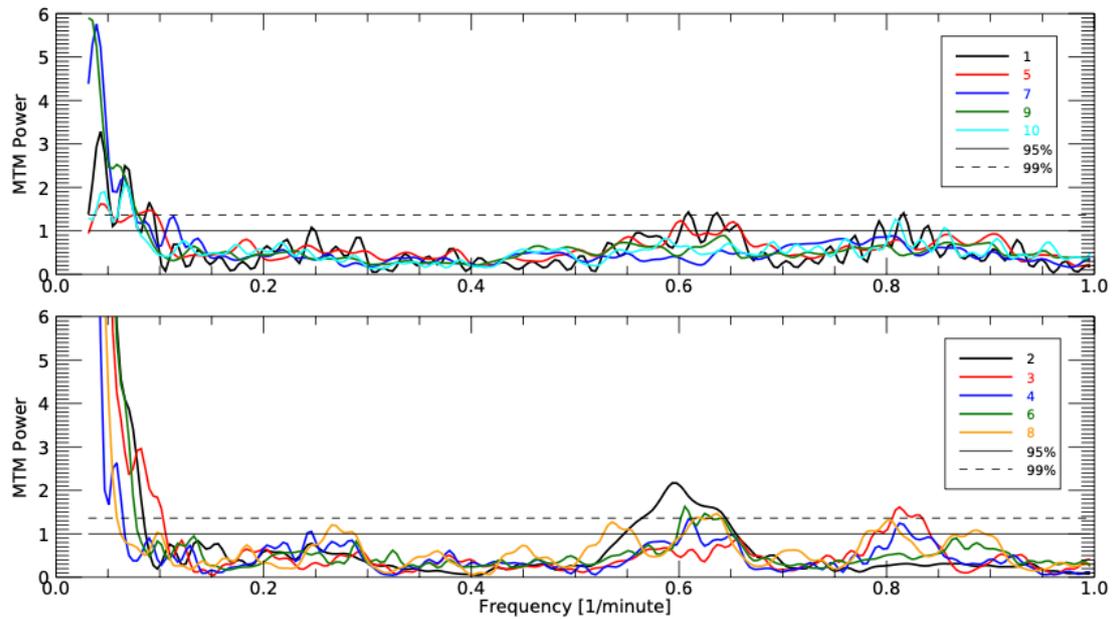

**Figure 9.** MTM period analysis results for the selected 10 longest living UDs. In the figure different UD presented in Figure 2 right panel and color shows power spectrum of this UD. Horizontal solid and dashed lines show 95 % and 99 % confidence levels, respectively.

**Table 1.** Maximum (Max), Mean and Minimum (Min) values of UDs characteristic

|      | Brightness | Diameter [km] | Eccentricity | Lifetime [min] | Dynamic Velocity [km/s] | Linear Velocity [km/s] |
|------|------------|---------------|--------------|----------------|-------------------------|------------------------|
| Max  | 783.08     | 275.55        | 0.65         | 65.75          | 3.84                    | 3.05                   |
| Mean | 145.54     | 206.56        | 0.29         | 6.92           | 0.76                    | 0.34                   |
| Min  | 12.74      | 107.75        | 0.02         | 0.75           | 0.05                    | 0.01                   |